\documentclass[doublecol]{epl2}
\usepackage{graphicx}% Include figure files
\usepackage{dcolumn}% Align table columns on decimal point
\usepackage{bm}% bold math

\begin{document}

\title{Spin wave free spectrum and magnetic field gradient of\\ nanopatterned planes of ferromagnetic cobalt nanoparticles:\\ key properties for magnetic resonance based quantum computing.}

\author{K. Benzid\thanks{$\:$ 1-2/ benzid@unistra.fr}, D. Muller\thanks{$\:$ 3/ dominique.muller@icube.unistra.fr}, P. Turek\thanks{$\:$ 1/ turek@unistra.fr}, J. Tribollet\thanks{$\:$ 1/ tribollet@unistra.fr : corresponding author}}               

\institute{1/ Institut de Chimie (UMR 7177 CNRS-UDS), $Universit\acute{e}$ de Strasbourg,\\ 4 rue Blaise pascal, CS 90032, 67081 Strasbourg cedex, France\\\\2/ $D\acute{e}partement$ de physique, Laboratoire de Physique Quantique et $Syst\grave{e}mes$ Dynamiques,\\ $Universit\acute{e}$ de Ferhat Abbas $S\acute{e}tif$ 1, $Alg\acute{e}rie$\\\\3/ Laboratoire ICube (UMR 7357 CNRS-UDS), $Universit\acute{e}$ de Strasbourg,\\ 23 rue du Loess, BP 20, 67 037 Strasbourg cedex 2, France.}

\abstract{We present a study by ferromagnetic resonance at microwave Q band of two sheets of cobalt nanoparticles obtained by annealing $SiO_{2}$ layers implanted with cobalt ions. This experimental study is performed as a function of the applied magnetic field orientation, temperature, and dose of implanted cobalt ions. We demonstrate that each of those magnetic sheet of cobalt nanoparticles can be well modelled by a nearly two dimensional ferromagnetic sheet having a reduced effective saturation magnetization, compared to a regular thin film of cobalt. The nanoparticles are found superparamagnetic above around 210 K and ferromagnetic below this blocking temperature. Magnetostatic calculations show that a strong magnetic field gradient of around 0.1 G/nm could be produced by a ferromagnetic nanostripe patterned in such magnetic sheet of cobalt nanoparticles. Such a strong magnetic field gradient combined with electron paramagnetic resonance may be relevant for implementing an intermediate scale quantum computer based on arrays of coupled electron spins, as previously reported (Eur. Phys. J. B (2014) 87, 183). However, this new approach only works if no additional spin decoherence is introduced by the spin waves exitations of the ferromagnetic nanostructure. We thus suggest theoretically some possible magnetic anisotropy engineering of cobalt nanoparticles that could allow to suppress the spin qubit decoherence induced by the unwanted collective excitation of their spins.}       


\maketitle

Magnetic nano-objects have many potential applications. Magnetic nanoparticles (NPs) can be used as contrats agent in the diagnosis and treatment of cancer~\cite{Gallo2013}, magnetic nanostripes can be used as a medium for efficient classical data transmission and processing~\cite{Khitun2010}, and magnetic nanodots can be used as storage elements for high density magnetic data recording~\cite{Thompson2000}. The present work reports on a new potential application of metal magnetic nanoparticles embedded in dielectric matrix. This is related to quantum information processing and electron paramagnetic resonance (EPR) spectroscopy. It was recently theoretically demonstrated~\cite{tribolletQCgradB2014} that the strong magnetic field gradient produced by a ferromagnetic nanostripe combined with the microwave pulses delivered by a pulsed electron paramagnetic resonance spectrometer, could constitute two of the three key elements constituting the hardware of a potential small scale spin based quantum computer, the third key element being the coupled electron spin qubits themselves placed nearby the ferromagnetic nanostripe. In this previous work~\cite{tribolletQCgradB2014}, it was also shown that unwanted spin decoherence could occur during information processing by microwave pulses if the paramagnetic resonances of the electron spin qubits spectrally overlapp the spin wave resonances existing in a bulk like ferromagnetic nanostripe. The solution to this problem that was proposed consist in a carefull design of the spin wave spectrum expected for a bulk like ferromagnetic nanostripe. This is a feasible but rather complex approach. Here, we suggest an alternate approach to achieve the same objective. This consist in the patterning of one or several ferromagnetic nanostripes in a thin film containing magnetic metal nanoparticles. One then expects a strong magnetic field gradient produced by such a diluted magnetic nanostripe, although it should be somehow reduced as compared to the one produced by a bulk like ferromagnetic nanostripe. However, such  magnetic dilution may be an advantage for this new particular kind of application according to the superparamagnetic behavior of such features. The solely strong spin wave mode expected at relatively low microwave frequencies (X or Q bands) is the uniform mode of precession of magnetization, generally called the ferromagnetic resonance mode in the case of a bulk ferromagnet. Avoiding the spectral overlap problem in this context thus corresponds to shifting this usually broad ferromagnetic-like resonance mode, sufficiently far from the paramagnetic resonances of the qubits. As it will be shown here, combining experiments and theory, this could be done by an appropriate engineering of the magnetic anisotropy of each magnetic metal nanoparticle present inside the diluted ferromagnetic nanostripe.\\\\In the first part of the article we present FMR experiments, similar to previous ones performed at X band~\cite{NPCopreviousFMR}, but here performed at microwave Q band and as a function of the direction of the applied magnetic field, on two cobalt nanoparticles sheets obtained by implantation of cobalt ions in a dielectric matrix of $SiO_{2}$, followed by thermal annealing. This FMR experimental study allow us to demonstrate that each of those magnetic sheets of cobalt nanoparticles can be fairly modelled by a nearly two dimensional ferromagnetic plane having a reduced effective saturation magnetization, with respect to a true bulk like thin film of cobalt. Subsequent calculations based on magnetostatics then demonstrates that a strong magnetic field gradient of around 0.1 G/nm can be produced by such a magnetic nanostripe made of cobalt nanoparticles. Since the key issue of the present work is quantum information processing of electron spin qubits, we show that a suitable engineering of the magnetic anisotropy of the cobalt nanoparticles may allow the individual microwave adressing and coherent manipulation of spin qubits state through EPR. This is done upon separating the spin resonance of the nanostripes from the spin resonances of the qubits.\\
\begin{figure*}[htbp]
\centering \includegraphics[width=0.6\textwidth]{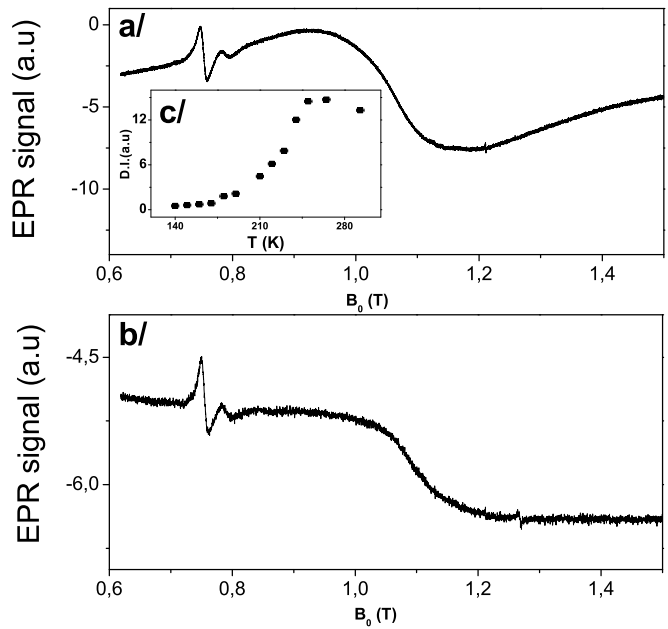}
\caption{\label{fig_01} Ferromagnetic resonance (FMR) spectrum obtained at Q band (33.95 GHz) with the magnetic field applied in the plane ($0^{o}$) of the cobalt magnetic nanoparticles, (1a) for sample I ($1.\!\;10^{17}\:cm^{-2}$ $Co^{+}$ ions) and (1b) for sample II ($0.5\!\;10^{17}\:cm^{-2}$ $Co^{+}$ ions). The inset (1c) inside figure 1a shows the temperature dependence of the double integration (noted D.I.) of the FMR signal which is proprotionnal to the effective magnetization of the cobalt nanoparticles detected by FMR spectroscopy at Q band. The modulation frequency was 100 kHz. The modulation amplitude was 4 G. The microwave power used was 2 mW. The narrow linewidth signals seen on the spectrum of the two samples below 0.8 Tesla are due to impurities in the Q band microwave cavity used.}  
\end{figure*} 
Two amorphous $SiO_{2}$ thin films, obtained by oxidation of Si wafers, were implanted at room temperature (295 K) with 160 keV $Co^{+}$ ions at two different doses, $0.5\!\;10^{17}\:cm^{-2}$ (sample I) and $1.\!\;10^{17}\:cm^{-2}$ (sample II), and further annealed at 873 K under hydrogen flow, in order to produce the metallic cobalt nanoparticles inside the $SiO_{2}$ matrix. Similar samples implanted at 160 keV with a nominal dose of $1.\!\;10^{17}\:cm^{-2}$ $Co^{+}$ ions have been previously investigated by TEM and Squid magnetometry~\cite{DMuller2001}. This previous study has shown that the magnetic cobalt metal nanoparticles have mainly the hexagonal phase, with a c axis of magneto-cristalline anisotropy randomly oriented in the amorphous $SiO_{2}$ matrix, that the average diameter of those magnetic cobalt nanoparticles is around 4.5 nm, and that they are superparamagnetic at room temperature, as shown by Squid magnetometry. The ferromagnetic resonance study of sample I and II containing the magnetic cobalt nanoparticles is performed using an EMX Bruker continuous wave electron paramagnetic resonance spectrometer operating at Q band (microwave frequency around 34 GHz). Modulation coils are used to modulate the EPR signal at 100 kHz, which allows to perform a sensitive lock in detection of the microwave absorption signal. This modulation also produces spectrum which appear as the derivative of a standard microwave absorption spectrum with gaussian or lorentzian lines. The cobalt implanted thin films of $SiO_{2}$ can be rotated inside the microwave resonator in order to vary the direction of the applied static magnetic field, from in plane (angle: $0^{o}$) to out of plane (angle: $90^{o}$). The sample temperature can also be varied from 4 K to 300 K using an Oxford Helium flow cryostat with a temperature controller designed for EPR experiments. \\\\ The FMR spectra obtained for sample I and II are shown on figure 1. The FMR resonance at Q band for a magnetic field applied in the plane ($0^{o}$) occurs at around 1,06 T for sample I and at around 1,09 T for sample II. As expected, due to the random orientation of the c axis of the hexagonal cobalt nanoparticles and also due to their size distribution resulting from the implantation process, the FMR signal is broad (pic to pic linewidth of around 0.2 T) for the two samples. Also, the double integration of the FMR signal of sample I is larger than the one of sample II, as expected given the larger ion dose in this sample. The in plane ($0^{o}$) magnetization of sample I has also been recorded as a function of temperature (inset 1c of figure 1a). As it was previously shown by SQUID magnetometry~\cite{DMuller2001} on similar samples, the cobalt nanoparticles of this sample are superparamagnetic at room temperature and ferromagnetic at low temperature. We found a transition between the two regimes occuring around a blocking temperature of around $T_{B}\:=\:210\:K$. Experimentally, the sample was cooled below 100 K without any magnetic field applied. As the applied magnetic field is set to zero between two successive steps, one expects that during FMR experiments performed below $T_{B}$, the magnetization of each nanoparticle is locked along the hexagonal c axis, which is random over the ensemble of nanoparticles. When the FMR spectrum is acquiered, the magnetic field rapidly increases, but it does not produce any alignement of the magnetization vectors of the nanoparticles due to the too slow magnetization relaxation dynamics below the blocking temperature. For temperatures far below $T_{B}\:=\:210\:K$, this slow relaxation dynamics leads to an almost zero magnetization signal measured over the ensemble of cobalt nanoparticles with randomly oriented c axis. Around $T_{B}\:=\:210\:K$ and above, the relaxation time of the magnetic nanoparticles of cobalt becomes much faster than the time scale requiered for the full FMR field sweep spectrum acquisition. As a consequence, around $T_{B}\:=\:210\:K$ and above, the magnetic nanoparticle start to align along the direction of the applied magnetic field producing a net magnetization.
\begin{figure*}[t]
\centering \includegraphics[width=1.0\textwidth]{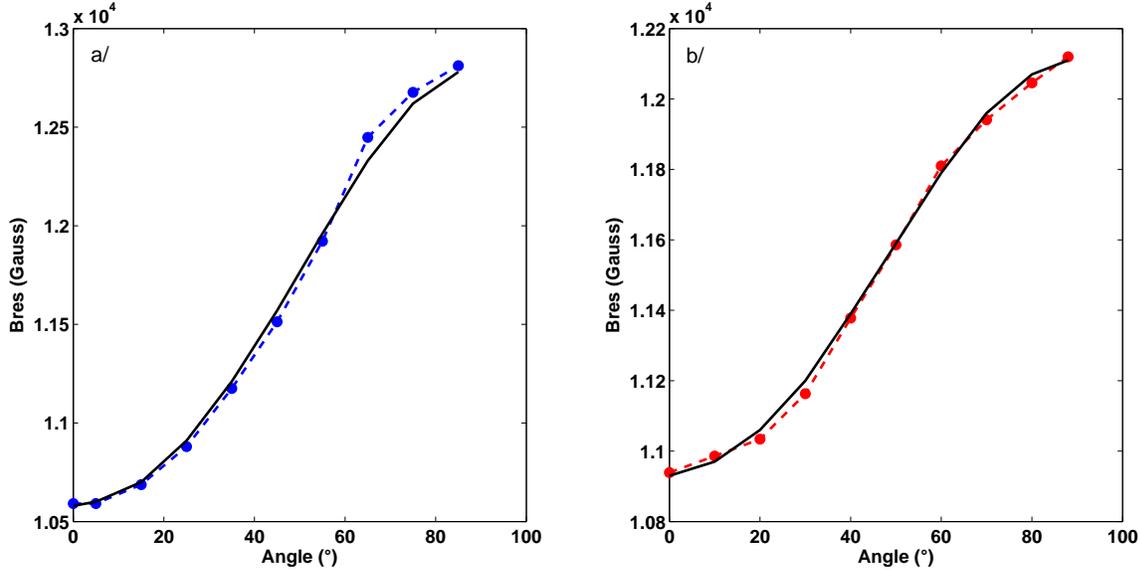}
\caption{\label{fig_02} Full rotational pattern of the ferromagnetic resonance mode of sample I (a/), and of sample II(b/), measured at room temperature and at Q band. Solid lines are numerical simulation obtained using the Smit and Beljers formalism (see text).}
\end{figure*}
The theoretical magnetic relaxation time $\tau_{1}$ of an isolated superparamagnetic nanoparticle is given by a standard Arrhenius law expression~\cite{Madsen2008}. It takes into account the temperature T and the magnetic anisotropy energy $K\:V$ of the nanoparticle of volume V. It is given by  $\tau_{1}\:=\:\tau_{0}\:exp\left(\frac{K\:V\:}{k_{B\:}T}\right)$. The intrinsic time scale $\tau_{0}$ is generally in the range of $10^{-9}\:s\:\leq\:\tau_{0}\:\leq\:10^{-13}\:s$. Here, the measurement time considered is the time requiered for the acquisition of a full field sweep FMR spectrum over more than 8000 Gauss, which is around $\tau_{mes}\:=\:100\:s$ here. This measurement time roughly correspond to the time window let to the nanoparticle to overcome the magnetic anisotropy barrier and thus to reorient itself in the direction of the applied magnetic field at the temperature T. Taking $\tau_{0}\:=\:10^{-9}\:s$, as in the previous analysis of similar samples investigated by SQUID magnetometry~\cite{DMuller2001}, and using $\tau_{mes}\:=\:100\:s$ for the FMR experiments, one obtains the following relation between the magnetic anisotropy energy $K\:V$ of the nanoparticle of volume V, and its blocking temperature $T_{B}$ (blocking energy $k_{B\:}T_{B}$ ): $K\:V\:\approx\:25\:k_{B\:}T_{B}$. Using the average diameter of $R\:=\:4.5\:nm$ obtained by TEM analysis of similar samples implanted by 160 keV cobalt ions at a dose of $1.\!\;10^{17}\:cm^{-2}$~\cite{DMuller2001}, and using the standard value of the constant of anisotropy for cobalt, $K\:=\:2\:10^{6}\:erg\:cm^{-3}$ ~\cite{Jamet2001}, one estimates $T_{B}\:=\:218 K$. This shows that our experimental measurements by FMR spectroscopy are in very good agreement with well known cobalt properties and with previous datas obtained by TEM and Squid magnetometry on similar samples.\\Now, using the standard Kittel formula\cite{Kittel1948} for the FMR of a thin ferromagnetic film with an in plane applied magnetic field, it is possible to estimate the effective saturation magnetizations of those thin films of cobalt nanoparticles. However, it is more reliable to estimate the effective saturation magnetizations of the films by recording the full rotational patterns of their ferromagnetic resonance mode, as it is shown on figure 2a and 2b respectively for sample I and II. As a matter of fact, the observed rotational pattern symmetry is due here to the shape anisotropy of each magnetic film, which is related to its effective saturation magnetization. Numerical simulations of the rotational patterns, performed using the Smit and Beljers formalism\cite{SmitandBeljers1955} and taking into account shape anisotropy, thus confirms that the dipolar couplings between the cobalt nanoparticles can not be neglected in those two magnetic films. From this point of view, the two films of magnetic nanoparticles can thus be seen as mimicking bulk like magnetic thin films with a reduced effective saturation magnetization compared to the bulk material ($B_{sat,\!bulk}\:\approx\:1.8 4\:T$). The results of the simulation made for sample I gives $B_{sat,\:I}\:\approx\:0.15\:T$ and the one made for sample II gives $B_{sat,\:II}\:\approx\:0.079\:G$. The ratio of the effective saturation magnetizations of the two films is very close to 2. It nicely corresponds to the ratio of the dose of implanted Co ions between sample I and sample II. Also, one notes that the ratio between the effective saturation magnetization of each film and the bulk saturation magnetization, noted f, is given by $f_{I}\:=\:\frac{0.15}{1.84}\:\approx\:0.081$ for sample I and $f_{II}\:=\:\frac{0.079}{1.84}\:\approx\:0.043$ for sample II. This ratio f should correspond to the averaged atomic fraction of cobalt in the $SiO_{2}$ matrix, as it can be measured by Rutherford BackScattering experiments (RBS)\cite{PhDORLEANS} . Those RBS measurements were previously performed\cite{PhDORLEANS} on similar samples obtained by implantation of $Co^{+}$ ions at 160 keV with a dose of $1.\!\;10^{17}\:cm^{-2}$ and revealed an averaged atomic fraction of cobalt in the $SiO_{2}$ matrix of 
$f_{RBS, \:1.\!\;10^{17}\:cm^{-2}}\:\approx\:0.13$. The present FMR experiments give $f_{FMR, \:1.\!\;10^{17}\:cm^{-2}}\:\approx\:0.08$. This reduced value shows that part of the cobalt NPs are not metallic nanoparticles but instead oxidized nanoparticles. The ratio of the metallic cobalt amount over the total cobalt amount in the $SiO_{2}$ matrix, $R_{FMR/RBS}\:=\:\frac{f_{FMR}}{f_{RBS}}\:\approx\:0.61$, is in good agreement with the ratio $R_{Squid}\:\approx\:0.72$ that was previously determined by Squid magnetometry on similar samples~\cite{DMuller2001}.\\The suggested application of the present work to quantum computing requires a strong magnetic field gradient~\cite{tribolletQCgradB2014}. The reported FMR experiments demonstrate that the two films of magnetic nanoparticles can be modelled as effective continuous magnetic thin film. Therefore, to get a strong magnetic field gradient one may elaborate nanostructures, e.g. nanostripes, within such nano-implanted thin films. This could be done, as it is shown on figure 3a and 3b, by an etching process, following either a deep UV optical lithography or electron beam lithography processing of a resist, depending on the targeted nanostripe dimensions. One such isolated and magnetically diluted ferromagnetic nanostripe of cobalt nanoparticles could thus produce the required strong magnetic field gradient as shown in figure 4 by the theoretical simulations based on magnetostatic calculations~\cite{tribolletQCgradB2014}. One has assumed in figure 4 a diluted ferromagnetic nanostripe of cobalt nanoparticles, having a width W= 1000 nm along the z axis, a thickness of T= 100 nm, corresponding roughly to the halfwidth of the cobalt ion implantation profile in the $SiO_{2}$ matrix\cite{PhDORLEANS}, and an infinite length along the y axis (100 microns). For the diluted ferromagnetic nanostripe of cobalt nanoparticles considered here, the effective saturation magnetization is $M_{sat,\:eff}$, with $\:\mu_{0}\:M_{sat,\:eff}\:=\:B_{sat,\:eff}\:=\:1840\:G\:=\:\frac{M_{sat,\:bulk Co}}{10}$. The figure 4a plots the dipolar magnetic field produced by this nanostripe, $B_{z}\left(z,\:x_{optim}\:=\:290\:n\:m\right)$, at the optimal position $x_{optim}$ for the qubits discussed below (see figure 3 for the definition of x and z axis) and as a function of the in plane position z of the qubit. Figure 4a thus demonstrate the good in plane homogeneity of the dipolar magnetic field gradient produced by the nanostripe in a plane placed at the distance $x_{optim}\:=\:290\:n\:m$ above (or below) this nanostripe and for z values close to zero. This in plane homogeneity is investigated more quantitatively on figure 4b using the homogeneity coefficient $C\left(x\right)$ previously introduced, $C\left(x\right)\:=\:\int^{+100}_{-100}\:dz\:\left(B_{z}\left(x,\:z\right)\:-\:B_{z}\left(x,\:0\right)\right)$ (see~\cite{tribolletQCgradB2014}). One defines the  position $x_{optim}$ by $C\left(x_{optim}\right)\:=\:0$. For the diluted ferromagnetic nanostripe considered here, one finds $x_{optim}\:=\:290\:n\:m$. The figure 4c then plots $B_{z}\left(z\:=\:0,\:x\right)$, which allows to determine the shift of the resonant magnetic field of the paramagnetic qubits for a given position x. Finally, the figure 4d shows the gradient along x of the dipolar magnetic field produced by the ferromagnetic nanostripe, $\frac{d\:B_{z}\left(x\right)}{dx}$ (assuming z=0 and y=0). Figure 4d shows that one could obtain a strong magnetic field gradient having a maximum strength of more than 0.1 G/nm at the optimal position $x_{optim}\:=\:290\:n\:m$ above this nanostripe.
\begin{figure*}[htbp]                                          
\centering \includegraphics[width=0.6 \textwidth]{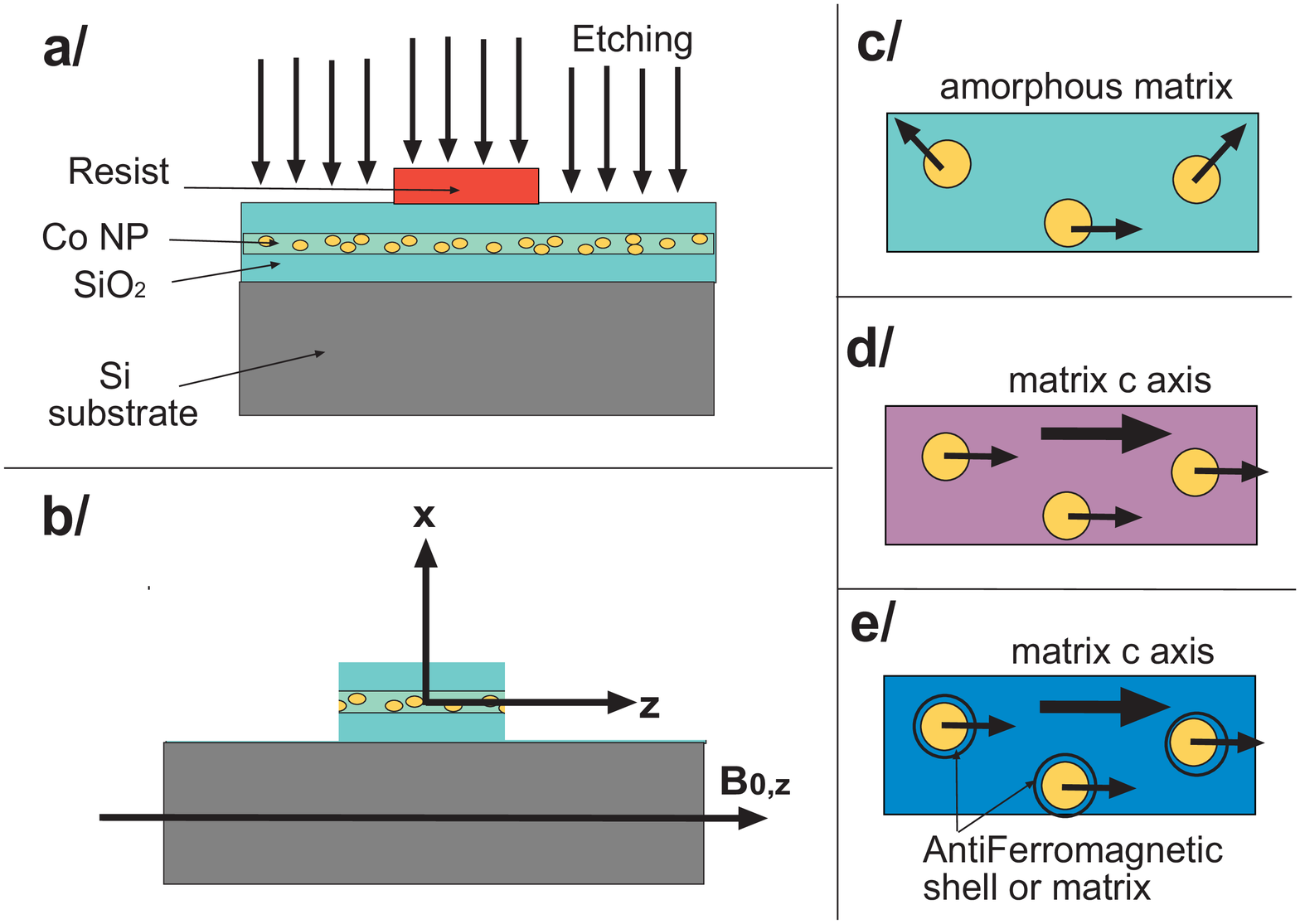}
\caption{\label{fig_03} 3a: Nanofabrication process, based on etching through a mask, suggested to produce the diluted ferromagnetic nanostripe discussed in the text and shown on figure 3b. Figures 3c, 3d and 3e illustrate the physical concepts behind the magnetic anistropy engineering of the nanoparticles discussed in the text.}  
\end{figure*}
 This position is also the one where this nearly one dimensionnal magnetic field gradient has its maximum in plane homogeneity, as it was discussed above. $x_{optim}$ is thus the position where many electron spin qubits should be placed nearby the nanostripe for quantum computing~\cite{tribolletQCgradB2014}. Note that the dipolar magnetic shift of the paramagnetic resonance lines of the qubits is expected to be much smaller at $x_{optim}$ in the case of such diluted ferromagnetic nanostripe of cobalt nanoparticles, than in the case of a true bulk like nanostripe (see figure 4c). As a consequence, and also due to the broad ferromagnetic resonance mode of the cobalt nanoparticles a spectral overlap is expected between the broad FMR mode of cobalt nanoparticles and the weakly shifted paramagnetic resonances of the electron spin qubits of the quantum nanodevice (g=2.00 is assumed here for the qubits). This may be adressed by a careful engineering of the magnetic anisotropy of the cobalt nanoparticules, as illustrated on figure 3c, 3d and 3e. The optimal strategy is illustrated by figure 3e.
\begin{figure*} [t] 
\centering \includegraphics[width=1.0\textwidth]{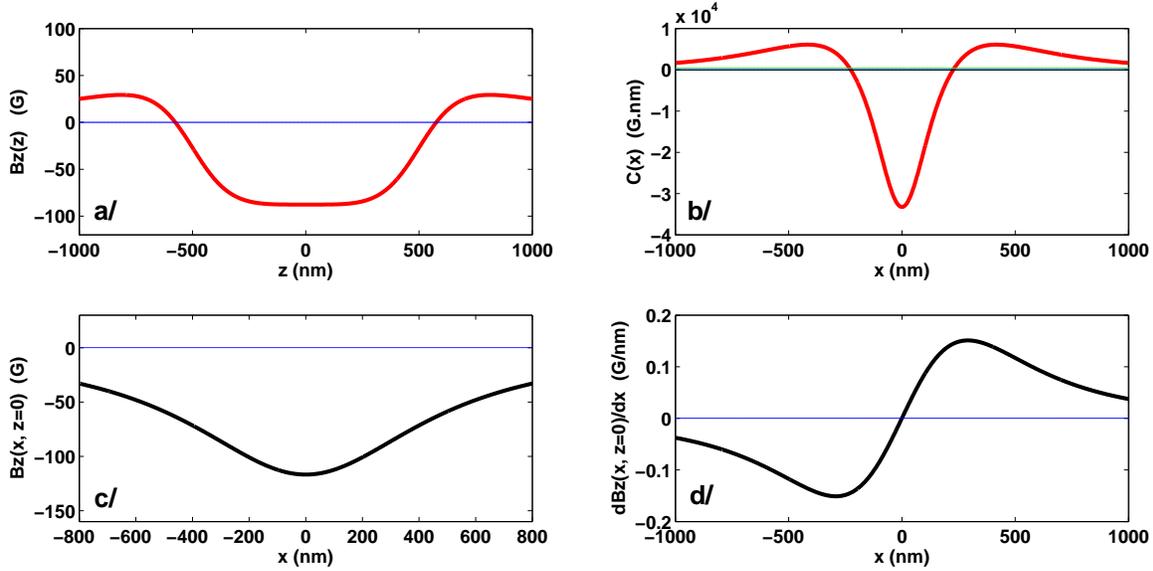}
\caption{\label{fig_04} Magnetostatic properties of the dipolar magnetic field produced by a diluted ferromagnetic nanostripe of cobalt nanoparticles. See text for details concerning the parameters used for the calculations.}
\end{figure*}
Figure 3c shows hexagonal cobalt nanoparticles with randomly oriented c axis in the amorphous $SiO_{2}$ matrix (this work and ~\cite{NPCopreviousFMR}).
\begin{figure*} [t]
\centering \includegraphics[width=1.0\textwidth]{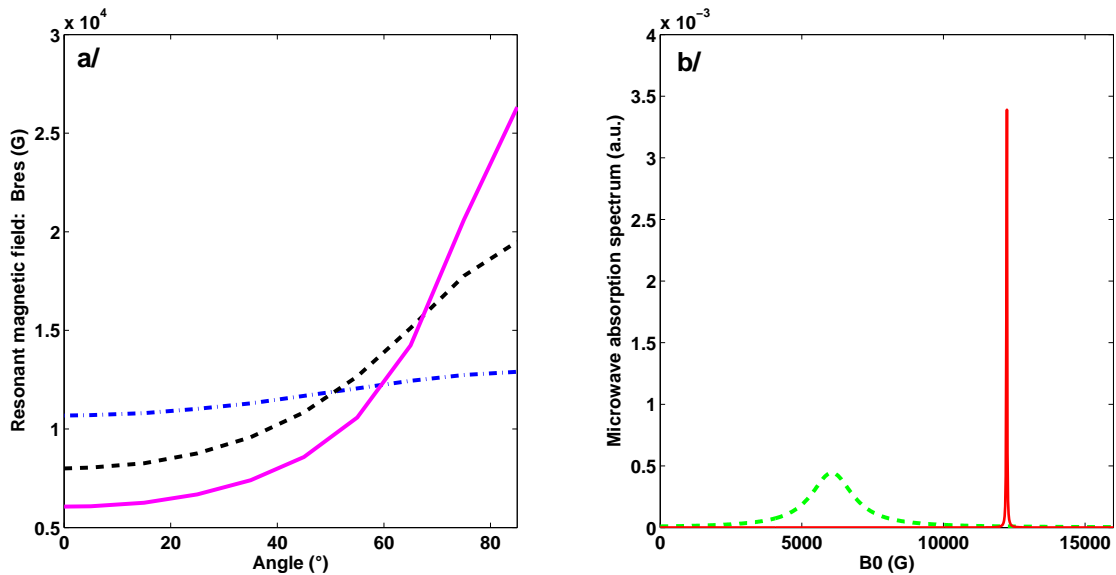}
\caption{\label{fig_05} 5a/ The three rotational patterns of the uniform FMR-like mode of the diluted superparamagnetic nanostripe expected in the three situations described qualitatively on figure 3c (blue dots: nanostripe with shape anisotropy alone), 3d (dark dots: nanostripe with shape anisotropy and with magneto-cristalline anisotropy of hexagonal cobalt with an effective anisotropy field $B_{A\!,\:MC}\:=\:\:\:6826\:G$, obtained from the K value in~\cite{Jamet2001}) and 3e (continuous violet: nanostripe with shape anisotropy , with magneto-cristalline anisotropy of hexagonal cobalt with the effective anisotropy field $B_{A\!,\:MC}\:=\:6826\:G$, and with exchange bias anisotropy due to the antiferromagnetic shells, with an effective anisotropy field $B_{A\!,\:EB}\:=\:7400\:G$, see~\cite{Nature2003}). 5b/ Full electron spin resonance spectrum expected for the hybrid paramagnetic qubits-superparamagnetic nanoparticles nanodevice. This hybrid nanodevice is assumed to contain electron spin qubits placed at a distance $x_{optim}\:=\:290\:n\:m$ above or below the diluted superparamagnetic nanostripe of cobalt NPs described on figure 3e. The linewidth of the uniform FMR-like mode was assumed to be around 1000 Gauss, and the one of the paramagnetic resonance of the qubits around 10 Gauss. $\:\mu_{0}\:M_{sat,\:eff}\:=\:B_{sat,\:eff}\:=\:1840\:G$. }
\end{figure*} 
Figure 3d shows a similar assembly but with NPs having their hexagonal c axis oriented along the in plane c axis of an anisotropic cristalline  matrix, like $Al_{2\:}O_{3}$~\cite{Meldrum2003}. Figure 3e shows another assembly similar to the one of figure 3d but here with all NPs surrounded and exchange coupled to a shell of antiferromagnetic material, whose spins are also directed along the c axis common to the cobalt nanoparticles and to the anisotropic cristalline  matrix~\cite{Nature2003,Lee2006}. A controlled annealing of the sample under an oxygen atmosphere may allow to control the antiferromagnetic shell around the metallic cobalt nanoparticles and thus to increase their resulting in plane magnetic anisotropy. Ion implantation of cobalt ions directly inside an anisotropic antiferromagnetic cristalline matrix followed by the same nanofabrication process could be another way to produce this strongly anisotropic diluted ferromagnetic nanostripe with an in plane easy axis of magnetization. The increase of the uniaxial in plane magnetic anisotropy, from case 3c to case 3e, is well demonstrated on figure 5, which shows the rotationnal pattern of the uniform FMR-like mode of such superparamagnetic implanted nanostripes in those three cases. It is observed (figure 5a), that increasing the in plane uniaxial anisotropy shift to lower magnetic field values the uniform FMR-like mode of the diluted superparamagnetic nanostripe for a in plane ($0^{o}$) applied magnetic field. The paramagnetic resonance of the electron spin qubits placed at $x_{optim}$ above the nanostripe is quite insensitive to this magnetic anisotropy engineering, whereas the FMR-like mode is much shifted towards lower fields as previously discussed. The figure 5b summarizes the expected situation.\\ As previously discussed~\cite{tribolletQCgradB2014}, the inter- qubits distances inside spin chains have also to be well chosen. Given that here the magnetic field gradient produced by the diluted ferromagnetic nanostripe is reduced compared to the one produced by a bulk like ferromagnetic nanostripe, dipolar coupled electron spin qubits would thus have to be separated by distances larger than the optimal 3.5 nm distance previously estimated, in order to distinguish their paramagnetic resonance and to coherently manipulate them by microwave pulse sequences~\cite{tribolletQCgradB2014}. This would be possible but would then lead to a further reduced scalability of the small scale quantum computer nanodevice. To avoid this new problem occuring in the new nanodevice design proposed presently, one could use stronger spin-spin coupling between successive spin qubits inside the spin chains, like exchange couplings, thus keeping an interesting number of spin qubits in such small scale quantum processor~\cite{tribolletQCgradB2014}. ZnO bulk single crystals are particularly interesting anisotropic single crystals that could be tested for such purpose. The first reason is that ZnO has the wurtzite anisotropic structure (c axis of anisotropy) and that this kind of alignement of the anisotropy axis of cobalt NPs has been already observed~\cite{Norton2003}. The second reason is that ZnO is a semiconductor matrix with a very low spin orbit coupling and which can be isotopically purified. This implies that it is an excellent host matrix for electron spin qubits, like the shallow indium donors~\cite{tribolletIndonor}, or transition metal ions, like the $Fe^{3+}$ spin qubits~\cite{tribolletFe} and the $Mn^{2+}$ spin qubits~\cite{tribolletMn}, that we previously investigated by pulsed EPR spectroscopy. The third reason is that ZnO is a direct gap semiconductor with excellent optical properties, including a very large exciton binding energy. This could be used for producing an effective optically induced long range exchange couplings between the qubits present in the ZnO matrix~\cite{ORKKY1,ORKKY2}, as requiered here due to the reduced strength of the available magnetic field gradient produced by a superparamagnetic nanostripe.\\

In conclusion, we have demonstrated by a ferromagnetic resonance study at microwave Q band of two cobalt nanoparticles sheets obtained by implantation of cobalt ions in a dielectric matrix of $SiO_{2}$, followed by thermal annealing, that each of those magnetic sheets can be well modelled by a nearly two dimensional ferromagnetic plane having a reduced effective saturation magnetization with respect to a true bulk like thin film of cobalt. Magnetostatic calculations have then shown that a strong magnetic field gradient of around 0.1 G/nm could be produced by a ferromagnetic nanostripe patterned in such magnetic plane of cobalt nanoparticles. This strong magnetic field gradient combined with electron paramagnetic resonance may be usefull for implementing an intermediate scale quantum computer based on arrays of coupled electron spins(J. Tribollet, Eur. Phys. J. B (2014) 87, 183). This is possible as far as the magnetic anisotropy engineering of the cobalt nanoparticles allows to overcome the problem of the spectral overlap between the narrow shifted paramagnetic resonances of electron spin qubits and the broad uniform FMR-like mode of the diluted superparamagnetic nanostripe. This magnetic anisotropy engineering of the cobalt nanoparticles could thus suppress the spin qubit decoherence induced by the unwanted spin waves excitations, and thus represents an alternative solution to the previously proposed solution requiring a carefull design of the spin wave spectrum of a bulk like ferromagnetic nanostripe. The price to pay for using this new design is to chose spin-spin couplings between qubits which are stronger, over large nanometer scale distances, than the dipolar couplings previously proposed. We finally suggested that this new strategy for quantum computing may be particularly well suited for exciton-mediated exchange coupled electron spin qubits in wurtzite zinc oxide, individually and coherently manipulated by means of microwave pulses and of the strong magnetic field gradient produced by a superparamagnetic nanostripe made of implanted cobalt nanoparticles.\\

\end{document}